\documentclass[twocolumn,aps,pre,showpacs]{revtex4}

\usepackage{tabularx}
\usepackage[latin1]{inputenc}
\usepackage{graphicx}
\usepackage{latexsym}
\usepackage{amssymb}
\usepackage{amsmath}
\usepackage{color}

\usepackage{amssymb}

\begin{document}
\def\d{{\rm d}}
\def\ex{{\rm e}}
\def\im{{\rm i}}
\def\e{{\bf e}}
\def\w{{\bf w}}
\def\g{{\bf g}}
\def\u{{\bf u}}
\def\x{{\bf x}}
\def\A{{\bf A}}
\def\c{{\bf c}}
\def\W{{\bf W}}
\def\k{{\bf k}}
\def\R{{\bf R}}
\def\r{{\bf r}}
\def\z{{\bf z}}
\def\smalze{{\scriptscriptstyle{(0)}}}
\def\smalun{{\scriptscriptstyle{(1)}}}
\def\smalU{{\scriptscriptstyle{U}}}
\def\smalV{{\scriptscriptstyle{V}}}
\def\smalS{{\scriptscriptstyle{(S)}}}
\def\smalss{{\scriptscriptstyle{T<T_i}}}
\def\smagrt{{\scriptscriptstyle{T>T_i}}}
\def\latent{{\cal L}}
\def\Bowen{{\cal B}}
\def\Entrainment{{\cal E}}
\def\Deposition{{\cal D}}
\def\calx{{\cal X}}
\def\caly{{\cal Y}}
\def\beq{\begin{eqnarray}}
\def\eeq{\end{eqnarray}}


\title{Limit regimes of ice formation in turbulent supercooled water}
\author{Francesca De Santi$^1$, Piero Olla$^1$}
\thanks{Email address for correspondence: olla@dsf.unica.it}
\affiliation{$^1$ ISAC-CNR and INFN, Sez. Cagliari, I--09042 Monserrato, Italy.}

\begin{abstract}
A study of ice formation in stationary turbulent conditions
is carried out in various limit regimes of
crystal growth, supercooling and ice entrainment
at the water surface. 
Analytical expressions for the temperature, salinity and ice concentration mean profiles 
are provided, and the role of fluctuations in ice production 
is numerically quantified. Lower bounds on the ratio of sensible heat flux to
latent heat flux to the atmosphere are derived and their dependence on key parameters such 
as salt 
rejection
in freezing and ice entrainment in the water column is elucidated. 
\end{abstract}

\maketitle

\section{Introduction}
\label{Introduction}
Ice production in polar oceans usually occurs in the presence of turbulence and wave motions
induced by strong winds. This prevents the formation of a continuous ice layer at the sea
surface in the initial phases
(thin ice films, called nilas, are indeed observed in very calm conditions). 
A slurry of ice crystals, with a characteristic milky or greasy appearance, called grease ice,
is generated instead. The ice crystals, called frazil crystals or frazil ice, have diameters 
ranging from 0.01 up to $\sim 4$ mm and thickness from 1 to $100\ \mu$m 
\cite{martin81}. If the wind is sufficiently strong, the frazil ice
may be blown away, leaving the water surface exposed to the cold air, thus
enhancing the ice production and the heat transfer to the atmosphere
\cite{alam98}. Agglomeration 
of the frazil crystals first into pancake shaped 
objects (so-called pancake ice \cite{doble03})
and then into larger floes, leads eventually to the formation of a compact ice
layer.

Frazil ice is typically present in the Marginal Ice Zone, which is the
transition region between the open polar ocean and the continuous
ice that covers the central basin, but is also present under ice shelves, in polynyas
and in leads. Ice production is
accompanied by salt rejection, and is thus believed to play
an important role in stimulating convection in ocean waters.
Frazil ice may also contribute to transport of nutrients
and other trace elements entrained in its body \cite{eicken00,smedsrud01}.

Frazil ice formation is a complex phenomenon in which many physical processes
play a relevant role \cite{daly94a}. We can list the most important ones.
\begin{itemize}
\item
Small droplets and foam
are continuously lifted up from the water surface and freeze in contact of the 
cold air. When they
return to the water column they act as primary ice seeds. 
\item
If the upper layers of the ocean are sufficiently supercooled, ice crystals grow out of the 
seeds and reach size up to the millimeter range.
New seeds are generated through fragmentation induced by collisions with
other crystals (secondary nucleation).
\item
Part of the crystals are entrained by the turbulence and are transported down the column. 
Additional ice production may take place away from the surface if the
supercooling is sufficient. Field data indicate that underwater frazil ice
and significant supercooling in the water column may indeed be present down to 
depths of 5--50 m (see e.g. \cite{ito15}). 
\end{itemize}

Theoretical models for the growth and transport of
frazil ice have been developed over the years.
In the one-dimensional theory of Omstedt and Svensson \cite{omstedt84},
the upper ocean was modelled as a turbulent 
Ekman
layer with the sea-ice mixture treated as a
continuum. This model was improved in
\cite{omstedt85,svensson98} to take into account
the size spectrum of the crystals.
Ice production under ice shelves
was studied in \cite{jenkins95}, 
adopting a
Boussinesq-like approximation.
This latter study was later extended in 
\cite{smedsrud04} to account for the size spectrum of the frazil crystals. 
A detailed study of
the precipitation of the frazil crystals on the shelf was carried out in 
\cite{holland05}.
The dynamics of frazil ice was included in regional ocean models  
\cite{galton-fenzi12}.
Large Eddy Simulations (LES) were utilized in 
\cite{skyllingstad01}
to study the frazil ice dynamics under polar
ice covers and leads, but no account was taken of the
size distribution of the ice crystals. Only recently there
has been some attempt to include such information in LES of
frazil ice formation in open ocean \cite{matsumura15}.

Due to the complexity of the process, all these models necessarily rely on parametrization 
of small scale phenomena and on the introduction of empirical constants. In such
circumstances, it may be of some interest to 
look for limit regimes in which a reduced number of
parameters is at play and identification of key physical aspects is simpler. 
This is precisely the strategy adopted in the present paper.
The analysis will focus on the
constraints imposed by the conservation laws and the
thermodynamics of the process. The case of 
a homogeneous domain is examined first, 
analyzing the relative importance 
of salinity and heat release in ice formation in controlling supercooling.
The analysis shifts then to the
real problem, i.e. ice production in a water column that is mechanically forced 
and simultaneously cooled down at the top surface. 
Lower bounds on the ratio of sensible
to total heat flux to the atmosphere, valid in stationary conditions, are derived. 
Predictions on the depth of the supercooled region and on the depth reached by frazil ice
are provided.

\section{Ice production: budget equations}
\label{Ice production}
Let us start by considering ice formation at constant pressure 
in a thermally isolated, initially supercooled volume of salt water.
The water is stirred vigorously to maintain uniform
conditions and to
ensure that only small ice crystals are present.
The volume is taken small enough for complications
associated with differences between temperature and potential temperature,
and with the depth dependence of the freezing point, to be negligible.
\begin{table}[h]
\footnotesize
\caption{\label{table1}
Physical parameters for salt water at reference temperature
$T_B=-2.09{\rm\,^oC}$, and total salinity $S_R=35\,$g/kg.
}
\begin{tabularx}{\columnwidth}{XX}
\hline
\hline
$a_S\simeq 0.0565\ {\rm ^oC/(g/kg)}$
& haline lowering of freezing point.
\\
$a_z\simeq 7.61\cdot 10^{-4}\ {\rm ^oC/m}$
& lowering of freezing point with depth.
\\
$\alpha_T  \simeq 3.79\cdot 10^{-4} {\rm \ m/(^oC\,s^2)}$ 
& thermal buoyancy coefficient.
\\
$\alpha_S \simeq 7.6\cdot 10^{-3}{\rm \ m/((g/kg)\,s^2)}$ 
& haline buoyancy coefficient.
\\
$\alpha_C \simeq 1{\rm \ m/s^2}$ 
& ice buoyancy coefficient.
\\
$c_P  \simeq 3947\, {\rm J/(kg\,^oC)}$ 
&   water specific heat.
\\
$\latent  \simeq 3.35\cdot 10^5\, {\rm J/kg}$ 
&  specific latent heat of fusion.
\\
$\rho_w  \simeq 1030\, {\rm kg/m^3}$ 
& reference water density.
\\
$\rho_i \simeq 920\, {\rm kg/m^3}$ 
& ice density.
\\
\hline
\hline
\end{tabularx}
\end{table}
We know that a variation $\delta C$ in the volume
fraction of ice leads to a heat release per
unit volume in the liquid phase,
\beq
\delta Q^L=\latent\rho_i\delta C,
\label{delta Q}
\eeq
where $\latent$ is the latent heat of fusion per unit mass of ice
and $\rho_i$ is the ice density. This will produce a temperature
increase in the liquid phase, 
\beq
\delta T=(\hat\rho\latent/c_P)\delta C,
\label{delta T}
\eeq
where we have indicated with 
\beq
\hat\rho=\rho_i/\rho_w\simeq 0.89
\eeq
the density ratio of ice and water and $c_P$ is the water specific heat
(we consider a small $C$ regime such that the ice contribution to the 
ice capacity of the medium is negligible; we also neglect
the contribution to the heat flux from the temperature difference between the
liquid phase and the ice \cite{holland05}, as it is much smaller than that from the latent heat).

During freezing a fraction $\beta\approx 1$ of the salt that was 
in the water forming the crystals is expelled to the surroundings \cite{daly94a}.
The local salinity can be expressed as the sum of a reference salinity
$S_R=35\ {\rm g/kg}$ and a deviation $S$ that is expected in most situations
to be small.
A volume fraction increase $\delta C$ in the ice thus
corresponds to a decrease $\hat\rho\delta C$ of the liquid volume fraction,
and to a release of salt per unit volume  
\beq
\delta S=\hat\rho\beta S_R\delta C.
\label{delta S}
\eeq
The water freezing temperature  
decreases for increasing pressure (depth) and salinity content. 
For small deviations, we have a linear relation \cite{daly94a}
\beq
T_i=-a_SS+a_zz,
\label{T_i}
\eeq
where $-z$ is the depth and $T_i$ is the deviation of the freezing temperature from the
reference value $T_{iR}\simeq -2.09{\rm ^oC}$ at salinity $S_R$ and zero depth. 
Similarly, $T$ will indicate from now on the deviation of the water temperature with respect
to $T_{iR}$, i.e. the supercooling at salinity $S_R$ and zero depth.

The decrease of the freezing point from creation of a volume fraction 
$\delta C$ of ice is
\beq
\delta T_i=-\hat\rho a_S\beta S_R\delta C.
\label{delta T_i}
\eeq
The supercooling in a water volume with initial supercooling $T_0-T_{i0}$ and
no ice, will be therefore, after formation of an ice volume fraction $C$,
\beq
T-T_i=T_0-T_{i0}
+(a_S\hat S_R+\hat \latent)C
\label{T-T_i}
\eeq
where 
\beq
\hat S_R=\hat\rho\beta S_R\quad{\rm and}\quad 
\hat\latent=\hat\rho\latent/c_P.
\label{hat latent}
\eeq

From Table \ref{table1} we find $a_S\hat S_R\simeq 1.76\ {\rm ^oC}$
and $\hat\latent\simeq 75.5\ {\rm ^oC}$.
Warming from latent heat
release is more effective in destroying supercooling
than freezing point lowering by salinity increase.
This implies that ice production can be sustained longer more
effectively by cooling down the mass of water than by removing salt.

We can determine the ice volume fraction that can be generated starting
from given supercooled condition by equating to zero the final supercooling
$T-T_i$.  From Eq. (\ref{T-T_i}), we find 
\beq
C_{sat}=\frac{T_{i0}-T_0}{\hat\latent+a_S\hat S_R}.
\label{C_sat 0}
\eeq
For an initial supercooling $T_0-T_{i0}=-0.1\ {\rm ^oC}$ of the order
of what is observed in wave-tank experiments \cite{ushio93}, and
considered as maximum transient supercooling in several models 
(see e.g. \cite{svensson98,skyllingstad01,matsumura15}),
we would get an ice volume fraction at saturation $C_{sat}\simeq 0.0013$.
Is it big or is it small? For a monodisperse suspension of disks of aspect ratio 
$\epsilon$,
the maximum volume fraction compatible with random orientation of the disks 
is $C\sim\epsilon$. This is what would be obtained if each 
volume of the fluid of size $\sim R^3$, with $R$ the radius of the disks,  
contained on the average one disk. Higher volume fractions
could be achieved (maintaining random orientation) if smaller
crystals filled the gaps among the disks to form a mortar-like assembly. 
An estimate of the aspect ratio of typical frazil crystals
is $\epsilon\approx 1/50$ \cite{smedsrud04,holland05}, which would lead to a 
threshold volume fraction for grease ice $C_g\approx 1/50$.
Higher volume fractions would correspond to more compact ice mixtures,
with the transition to solid ice occurring at $C\approx 0.3$ \cite{maus12}.  
Field measurements suggest a typical volume fraction of grease ice
$C_g= 0.2-0.3$ \cite{bauer83,smedsrud06}.
Even the lower estimate $C_g\approx 1/50$ 
is an order of magnitude above the saturation concentration 
$C_{sat}$ predicted by Eq. (\ref{C_sat 0}) at a supercooling $T-T_i\approx -0.1\ {\rm ^oC}$.
Transport of heat and salinity away from the production region, together with accumulation
of the ice crystals, are therefore necessary for a grease ice layer to be established.
\section{Transport}
\label{Transport}
We follow \cite{omstedt84,skyllingstad01,holland05} and others, and coarse grain the dynamics
at a scale such that the frazil ice can be treated as a continuum, described locally by 
the  volume fraction field $C(\r,t)$, which is supposed small throughout the analysis
(for extension to a large $C$ regime, see e.g. \cite{feltham06}).
Momentum transport is described by the Navier-Stokes equation, which in the Boussinesq
approximation reads
\beq
(\partial_t+\u\cdot\nabla)\u
+(1/\rho_w)\nabla  P=\nu\nabla^2 \u
\nonumber
\\
+(\alpha_T T+\alpha_C C-\alpha_SS)\hat\z
\label{Navier-Stokes}
\eeq
(we take the reference system with origin at the water surface, and the vertical $z$-axis
upward directed).
The ice buoyancy coefficient can be expressed in terms of the ice density ratio as 
$\alpha_C=g(1-\hat\rho)\simeq 1\ \rm m/s^2$, where 
$g\simeq 9.8\ {\rm m/s^2}$ is the gravitational acceleration. The values of 
the other coefficients $\alpha_T$ and $\alpha_S$ are listed in Table \ref{table1}.
The kinematic viscosity of salt water is $\nu\simeq 1.95\cdot 10^{-6}\ {\rm m^2/s}$.

We can determine the contribution to buoyancy that would be produced by
a local increase $\delta C$ in the ice volume fraction. Using Eqs. 
(\ref{delta T}) and (\ref{delta S}) to estimate the local increments of temperature
and salinity: 
\beq
\frac{\alpha_T\delta T}{\alpha_C\delta C}\approx 
0.03,
\quad{\rm and}\quad
\frac{\alpha_S\delta S}{\alpha_C\delta C}\approx 0.24.
\label{relative buoyancy}
\eeq
This tells us that, 
while the dominant contribution to ice production saturation  comes from latent
heat release, the dominant contribution to convection comes directly from 
ice loading, followed by salinity.

The three fields $T$, $S$ and $C$ are governed by equations
(see e.g. \cite{omstedt84})
\beq
&&(\partial_t+\u\cdot\nabla) T=\kappa_T\nabla^2 T+\Pi_T,
\label{T equation}
\\
&&(\partial_t+\u\cdot\nabla) S=\kappa_S\nabla^2 S+\Pi_S,
\label{S equation}
\\
&&[\partial_t+(\u+\u_r)\cdot\nabla] C=\Pi_C,
\label{C equation}
\eeq
where $\Pi_T$, $\Pi_S$ and $\Pi_C$ are production terms whose form will be specified below.

\begin{figure}
\begin{center}
\includegraphics[width=0.65\columnwidth]{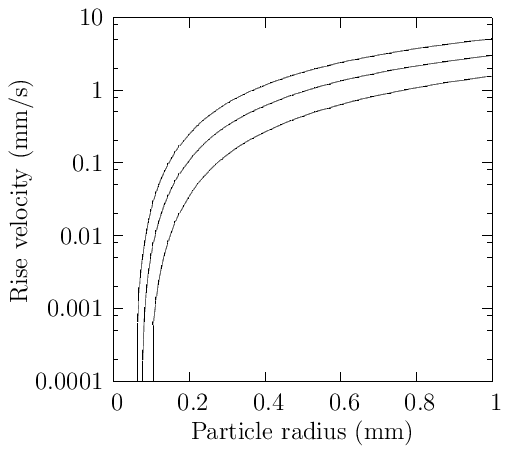}
\caption{Rise velocity $u_r$ as a function of the particle radius for different
values of the aspect ratio:
$\epsilon=1/100$ (bottom curve); $\epsilon=1/50$ (middle curve);
$\epsilon=1/30$ (top curve).
}
\label{frazfig1}
\end{center}
\end{figure}

The diffusivity coefficients in Eqs. 
(\ref{T equation}) and (\ref{S equation}) are
$\kappa_T\simeq 1.4\cdot 10^{-7}\ {\rm m^2/s}$,
and $\kappa_S\simeq 7.4\cdot 10^{-10}\ {\rm m^2/s}$,
and the molecular diffusivity of the ice crystals is disregarded.
Note that we have included in the equation for the ice fraction the rise velocity
of the crystals relative to the surrounding water, $\u_r=u_r\hat\z$. 
This is necessarily an average, as crystals of different size and aspect ratio will have
different rise velocity. Gosink and Osterkamp \cite{gosink83}
provided model equations for $u_r$, in the case of individual crystals, as a function of their 
radius and thickness.  A plot of such dependence is shown in Fig. \ref{frazfig1}.

The production terms can be taken in the form, from Eqs. (\ref{delta T}) and (\ref{delta S}):
\beq
\Pi_C=\Gamma C;
\quad
\Pi_S=\hat S_R\Gamma C;
\quad
\Pi_T=\hat\latent\Gamma C.
\label{Pi}
\eeq
Usually, a linear dependence of the growth rate $\Gamma$ on the supercooling is
assumed,
\beq
\Gamma=\gamma\,(T_i-T).
\label{Gamma}
\eeq
The parameter $\gamma$ depends on the size distribution and the morphology of the 
crystals \cite{fujioka74,hammar95,rees-jones15}. Some authors \cite{jenkins95,holland05}
hypothesize an asymmetry between melting and freezing rates, 
$\gamma_\smagrt\sim\epsilon^{-1}\gamma_\smalss$. In general $\gamma$ 
can be interpreted
as the inverse time, normalized to the supercooling, required by a crystal to reach mature size.
In the case of a mono-disperse distribution,
with aspect ratio of the crystals $\epsilon$, $\gamma$ can be evaluated, following
\cite{fujioka74,hammar95,rees-jones15}
\beq
\gamma\approx
\hat\gamma/R^2,
\qquad
\hat\gamma=\hat\gamma(\epsilon).
\label{gamma}
\eeq
Different authors provide different estimates for the parameter $\hat\gamma$. A widely used
approximation \cite{jenkins95,holland05,galton-fenzi12} is the one from \cite{hammar95}, 
$\hat\gamma\approx 3.7\cdot 10^{-9} \ {\rm m^2(^oC\,s)^{-1}}$;
a value higher by a factor $\sim (\epsilon\ln\epsilon)^{-1}$ is suggested in 
\cite{fujioka74,rees-jones15}, with experimental support for the latter choice 
in \cite{shimada97}. The two choices would give for frazil crystals of radius $R=1$ mm and
aspect ratio $\epsilon=50$, 
$\gamma\approx 0.0037\ {\rm (^oC\,s)^{-1}}$ and 
$\gamma\approx 0.04\ {\rm (^oC\,s)^{-1}}$, respectively.
\section{A stationary model}
\label{A stationary}
It is likely that a significant portion of grease ice forms during an initial transient in 
which supercooling is strong \cite{omstedt84}. After this, a quasi-stationary regime can be
expected, although this is necessarily an idealization, since external conditions (say weather 
patterns) vary on the same time scale of the process itself. An
approximation of statistical stationarity could 
nevertheless be used to describe the faster processes taking place near the
water surface. 

We envision a situation in which ice is formed primarily at the water surface. Part
of this ice may be blown away by the wind, 
and part of it accumulates at the surface
to form a grease ice layer, part of it is entrained in the water column. 
Additional ice formation may take place in the water column provided supercooling
is sufficient. 
The depth of the supercooled region 
will depend 
on the ratio between the removal rate of frazil ice by turbulence
and the rate of latent heat release
during ice growth. The depth reached by the frazil ice will depend 
on the turbulence intensity, the rise velocity of the crystals and
the depth of the supercooled region. 
Whether and where such additional ice forms, however, is difficult to ascertain 
in field observations \cite{ito15}. 

We consider the model situation of an infinitely deep, horizontally homogeneous water body
cooled from above in stationary conditions. We assume that no ice is present at sufficient
depth. This leads us to expect supercooling decreasing with depth both
at the top of the column and in the ice-free region. We use these assumptions as boundary
conditions in the ice production problem. 

In the analysis that follows we disregard the 
pressure contribution to $T_i$ (the $a_zz$ term in Eq. (\ref{T_i})). For a supercooling
scale $|T-T_i|\sim 0.1\ {\rm ^oC}$, this means restricting the analysis to the top few
tens of meters of the physical water column.

\begin{figure}
\begin{center}
\includegraphics[width=4cm]{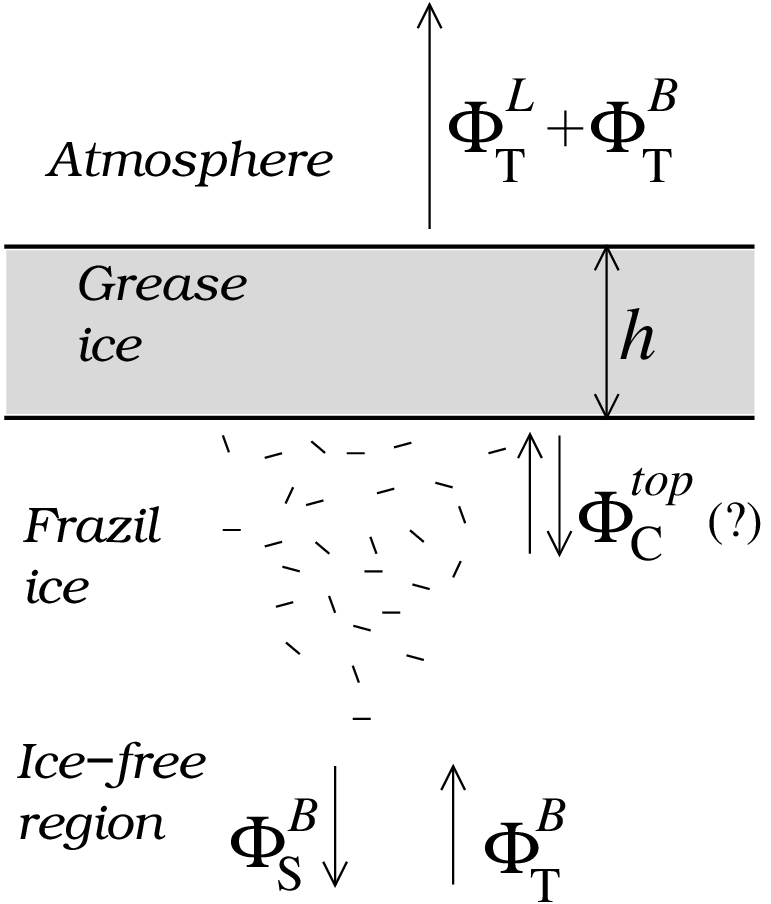}
\caption{Sketch of the heat, salinity and ice fluxes generated during build-up of
a grease ice layer. The heat flux to the atmosphere is split into latent heat $\Phi_T^L$
and sensible heat $\Phi_T^B$ contributions. The salinity flux $\Phi^B_S$ is downward directed.
The question mark indicates that the sign of the frazil ice flux $\Phi^{top}_C$ 
is {\it a priori} undetermined.
}
\label{frazfig2}
\end{center}
\end{figure}

Information on the way the heat, 
salinity and
frazil ice fluxes get organized in the domain can be obtained from simple budget considerations.
Ice formation leads to a downward directed salinity flux in the ice-free region, $\Phi_S^B$,
and to a latent heat contribution to the heat flux to the atmosphere $\Phi_T^L$.
A schematic of the processes is illustrated in Fig. \ref{frazfig2}.
%
%
%
%
From Eqs. (\ref{delta Q}) and (\ref{delta S}), we can express the latent heat flux in 
terms of the salinity flux, 
\beq
\Phi_T^L=-(\hat\latent/\hat S_R)\Phi^B_S
\label{Phi_ST^BL}
\eeq
(we express heat fluxes in units ${\rm ^oC\, m/s}$; conversion to natural units,
${\rm W/m^2}$, is achieved by multiplication with $\rho_wc_P$).
Putting together with the sensible heat flux $\Phi_T^B$, which coincides with the total
heat flux in the ice-free region, we get the total heat flux to the atmosphere
\beq
\Phi_T^{tot}=\Phi_T^B-(\hat\latent/\hat S_R)\Phi^B_S.
\label{Phi_T^tot}
\eeq
We can then introduce supercooling and neutral temperature fields
(we focus on a condition in which the pressure contribution to $T_i$ can be disregarded)
\beq
T_o=T+a_SS
\label{T_o}
\eeq
and
\beq
T_n=T-(\hat\latent/\hat S_R)S
\label{T_n}
\eeq
(note that $T_o<0$ in supercooled regions).
As can be checked from Eqs. (\ref{T equation}) and (\ref{S equation}), $T_o$ and
$T_n$ are decoupled, with only $T_o$ feeling the effect of ice formation and melting. The
flux $\Phi_{T_n}=\Phi_T-(\hat\latent/\hat S_R)\Phi_S$ is therefore conserved. We find
from Eq. (\ref{Phi_T^tot}),
\beq
\Phi_{T_n}=\Phi_T^{tot}.
\label{grande!}
\eeq
The temperature and salinity contributions to
buoyancy can be expressed in terms of the fields $T_{o,n}$: 
$T-(\alpha_S/\alpha_T)S\simeq{\rm const.}
-8.05\,T_o+9.05\,T_n$. We note that fixed $T_n$, a supercooled condition at large depth will
be less stable than one in which the bottom is above freezing.

The condition $\partial_z\bar T_o^B<0$ leads to the requirement that the supercooling
flux in the ice-free region is upward directed 
\cite{note00}
\beq
\Phi^B_{T_o}\equiv\Phi_T^B+a_S\Phi^B_S> 0.
\label{Phi_T^S}
\eeq
Equation (\ref{Phi_T^S}) tells us  that the ratio
$\Bowen=\Phi_T^B/\Phi_T^{tot}$ of sensible to total heat flux cannot be zero, the
minimum value being
\beq
\Bowen_{min}=
\frac{a_S\hat S_R}{\hat\latent+a_S\hat S_R}.
\label{Bowen}
\eeq
From Eq. (\ref{hat latent}), $\Bowen_{min}$ is an increasing function of the brine release 
coefficient $\beta$ and is maximum for $\beta=1$, $\Bowen_{min}\simeq 0.025$.
The smallness of $\Bowen_{min}$ has origin in the smallness of the contribution to supercooling
destruction from salinity release with respect to latent heat release in ice formation (see Eqs. 
(\ref{T-T_i}) and (\ref{hat latent}) and following discussion).
The maximum value $\Bowen=1$ describes a situation in which 
$\Phi_T^L=0$ and there is exact balance between ice production and melting.

\subsection{Frazil ice fluxes in the water column}
\label{The frazil}
Frazil ice in the water column acts as a sink term for the supercooling field $T_o$.
Equations (\ref{delta T}) and (\ref{delta S}) yields conservation of the flux
\beq
\Phi_\calx= \Phi_{T_o}-(\hat\latent+a_S\hat S_R)\Phi_C,
\label{Budget0}
\eeq
which is associated with the field 
\beq
\calx=T_o-(\hat\latent+a_S\hat S_R)C.
\eeq
A formula analogous to Eq. (\ref{grande!}) can be derived,
\beq
\Phi_\calx=\Phi_{T_o}^B.
\label{grande1}
\eeq
As in the case of $T_n$, the field $\calx$ is not affected by ice production.

The frazil ice flux $\Phi_C$ in the water column is the sum of the contributions by
deposition and turbulent entrainment,
\beq
\Phi_C=\bar Cu_r+\overline{\tilde u_z\tilde C}
\eeq
(overbar and tilde indicate average and fluctuating components). The 
flux on the top of the column, $\Phi^{top}_C$, will
be positive if deposition $\bar Cu_r$ dominates.
At stationarity,
positive $\Phi^{top}_C$ necessarily corresponds to ice production exceeding melting
in the water column.

Let us introduce the ratio $\Entrainment$ of the frazil flux at the top of the column,
taken with minus sign, $-\Phi_C^{top}$, and the total ice production $\Phi_T^L/\hat\latent$.
The two regimes $\Entrainment>0$ and $\Entrainment<0$ describe 
dominant entrainment and dominant deposition, respectively.
In terms of the parameter $\Bowen$,
\beq
\Phi_C^{top}=(\Entrainment/\hat\latent)(\Bowen-1)\Phi_T^{tot}.
\label{Entrainment}
\eeq
Note, from positive definiteness of the deposition flux, that we must
have $\Entrainment\le 1$. 

The assumption $\partial_z\bar T_o^{top}< 0$ 
allows us to refine the bound in Eq. (\ref{Bowen}).
From Eq. (\ref{Budget0}) and the condition that no ice is present at large depth, we
find 
\beq
\Phi^{top}_{T_o}=\Phi_{T_o}^B+(\hat\latent+a_S\hat S_R)\Phi_C^{top}>0.
\label{Budget}
\eeq
The salinity flux in the ice-free region,
$\Phi_S^B$, can be expressed in terms of the latent heat flux $\Phi_T^{tot}-\Phi_T^B=
\Phi_{T_n}-\Phi_T^B$: 
$\Phi_S^B=(\hat S_R/\latent)(\Phi_T^B-\Phi_{T_n})$. Substituting into the
supercooling and neutral fluxes in the ice-free region,
$\Phi^B_{T_o}=\Phi_T^B+a_S\Phi_S^B$ and
$\Phi_{T_n}=\Phi_T^B-(\hat\latent/\hat S_R)\Phi_S^B$, and eliminating
$\Phi_T^B$, allows us to write
\beq
\Phi_{T_n}=\frac{1-\Bowen_{min}}{\Bowen-\Bowen_{min}}\Phi_{T_o}^B,
\label{K_Bowen}
\eeq
where use has been made of Eqs. (\ref{grande!}) and (\ref{Bowen}).
Substituting into Eq. (\ref{Budget}) and exploiting Eq. (\ref{Entrainment})
finally gives
\beq
\Bowen>
\bar\Bowen_{min}(\Entrainment)=
\frac{\Bowen_{min}+\Entrainment}{1+\Entrainment}.
\label{refined bound}
\eeq  
In the limit 
$\Bowen\to\bar\Bowen_{min}(\Entrainment)$, all of the supercooling flux $\Phi_{T_o}^B$ is utilized
to melt the frazil ice in the column, and $\Phi_{T_o}^{top}=0$. 
For $\Entrainment=0$ (balance of entrainment and deposition),
we recover the bound in Eq. (\ref{Bowen}). The same situation occurs for $\Entrainment<0$
(deposition dominant with respect to entrainment), in which case the bound in Eq. (\ref{Bowen})
becomes stronger than the one in Eq. (\ref{refined bound}). In the limit case in which 
all the ice is entrained in the column, the sensible heat flux must be at least 
a fraction $\bar\Bowen_{min}(1)\simeq 1/2$ of the total heat flux.

\section{The grease ice layer}
\label{The grease}
In order for ice production to be maintained in the grease ice layer---if at all present---it 
is necessary that the salt released in the process is 
efficiently transported down the layer and dispersed in the water column. 
Due to the high viscosity of the medium ($\nu_g\simeq 0.01\ {\rm m^2/s}$ \cite{martin81a}),
turbulent transport is negligible. Wave induced random motions at the scale of the ice crystals,
however, are likely to enhance transport with respect to the case of a pure fluid
in analogous conditions. 
Some analytical progress is possible if we assume that this
is the dominant mechanism of transport, with idendical diffusivity
for and heat and salinity $\kappa_g\gg\kappa_{T,S}$.

The transport equations derived in Sec. \ref{Transport} can be extended to the finite
$C$ conditions characteristic of grease ice. We continue to assume stationary conditions,
in such a way that all the heat produced
during ice formation is transferred to the liquid phase and carried away by diffusion. 
Equations (\ref{T equation}) and (\ref{S equation}) become
\beq
&&\partial_z(\kappa_g\partial_z\bar T)+\hat\latent\Gamma \bar C=0,
\label{T equation grease}
\\
&&\partial_z(\kappa_g\partial_z\bar S)+\hat S_R\Gamma \bar C=0,
\label{S equation grease}
\eeq
where $\bar T$ and $\bar S$ 
refer to the liquid phase and a volume factor $1-\bar C$ is incorporated in $\kappa_g$.

The rate of ice formation is determined by the removal of heat and salt 
in the region around the crystals. We expect that the equations governing transport
at the microscale be linear in $T$ and $S$ \cite{note0}, so that for small $T_o$, 
$\Gamma\approx -\gamma_gT_o$, with $\gamma_g$ a constant dependent
on $C$ and on the crystal geometry.
We can obtain from Eqs. 
(\ref{T equation grease}) and (\ref{S equation grease}), equations for $T_o$ and $T_n$:
\beq
\partial_z(\kappa_g\partial_z\bar T_o)=\hat\latent\gamma_g\bar C\bar T_o
\quad{\rm and}\quad
\partial_z(\kappa_g\partial_z\bar T_n)=0.
\label{T_on equations}
\eeq
We identify in the first of Eq.  (\ref{T_on equations}) a characteristic length 
\beq
l\approx\Big(\frac{\kappa_g}{\gamma_g\hat\latent \bar C}\Big)^{1/2},
\label{l}
\eeq
which we allow to depend on $z$ on scale $h$.
We can make some estimates from parameters valid in the dilute regime.
Taking $\gamma_g\approx 0.04\ {\rm (^oC\,s)^{-1}}$, and $\bar C\approx 0.2$, 
we obtain $l$ values ranging from
millimeters (for $\kappa_g\approx\kappa_T$) 
to tens of centimeters (for
$\kappa_g\approx\nu_g\approx 0.01\,{\rm m/s^2}$).

Of the two relevant limits $l\ll h$ and $l\gg h$, only the first is of some interest.
For $l\gg h$, it is easy to see by Taylor expanding Eq. (\ref{T_on equations}), 
that ice production in the grease ice layer contributes to the heat flux with an $O(h/l)$ 
correction, so that, in the absence of ice formation in the water column, $\Bowen\simeq 1$.
The interest of the opposite limit regime $l\ll h$ lies in the fact 
that it allows an easier interpretation of
the lower bound in Eq. (\ref{Bowen}). Let us consider first the case
$\bar T(-h)=0$ (no ice formation or melting at the bottom of the layer). 

Equation (\ref{T_on equations}) has solution, 
for $z\gg -h$,
\beq
&&\bar T_o(z)=\bar T_o(0)\,\ex^{z/l_0}+{\cal T}(\ex^{-z/l}-\ex^{z/l_0}),
\label{T_o(z) grease}
\\
&&\bar T_n(z)=\bar T_n(0)+\bar T'_nz,
\label{T_n(z) grease}
\eeq
where $l_0\equiv l(0)$.
In the regime $l\ll h$, ${\cal T}\simeq-\bar T_o(0)\exp(-2\int_{-h}^0\d z/l(z))\simeq 0$, and
the second constant
$T'_n$ in Eq.  (\ref{T_n(z) grease}), is fixed by imposing that there is
no salt flux to the atmosphere, $\bar S'(0)=0$. This gives
\beq
\bar T'_n=\bar T_o(0)/l_0.
\label{T'_n}
\eeq
The fact that there is no supercooling at $-l\gg z\gg -h$, $\bar T_o(z)=0$, implies that 
temperature and salinity are dominated by $\bar T_n$ and have a linear profile
(see Eq. (\ref{T_n(z) grease})).
We can calculate the
heat and salinity fluxes $\Phi_T=-\kappa_g\bar T'$ and $\Phi_S=-\kappa_g\bar S'$ in this region. 
Substituting Eq. (\ref{T'_n}) and the condition $\bar T_o(z\ll-l)=0$ 
into Eqs. (\ref{T_o}) and (\ref{T_n}),
we get
\beq
&&\Phi_T=-\kappa_g\Bowen_{min}\bar T_o(0)/l_0;
\label{Phi_Tg}
\\
&&\Phi_S=(1-\Bowen_{min})\kappa_g\hat S_R\bar T_o(0)/(\hat\latent l_0).
\label{Phi_Sg}
\eeq
Since $\bar T_o(-h)=0$, there will be no melting at the bottom surface and the heat flux right
below will still be given by Eq. (\ref{Phi_Tg}). 
If the water in that region is
ice free, the heat flux will coincide with the sensible heat flux $\Phi^B_T$ and by
exploiting (\ref{T'_n}) and (\ref{grande!}) we obtain
$\Phi_T^B=\Bowen_{min}\Phi_T^{tot}$. Thus $\Bowen=\Bowen_{min}$ and the latent heat
contribution to the heat flux to the atmosphere is maximum.
A situation in which $\Bowen>\Bowen_{min}$ corresponds to
ice melting at the bottom of the layer, with  the
difference $(\Bowen-\Bowen_{min})\Phi_T^{tot}$ giving precisely the heat required for melting.
The same situation would arise if $\bar T_o$ remained close to zero at $z=-h$,
but frazil ice were entrained by turbulence and melted down upon reaching
the region with $\bar T_o>0$ at the bottom of the column.
The interpretation of $(\Bowen-\Bowen_{min})\Phi_T^{tot}$ as a melting heat is clearly
lost if the grease ice is so thin that it loses its insulating properties and the
excess heat is transferred to the atmosphere.
\section{The water column}
\label{The turbulent}
Frazil ice production requires strong winds. A minimum wind velocity 
$u_{wind}=4.35$ m/s at 10 m above the water surface was suggested in \cite{bauer83}
as a necessary condition for frazil ice formation. A turbulent boundary layer forced
both by mechanical stress and convection induced by heat and salinity fluxes is
thus expected to exist below the grease ice layer. 

An estimate of the friction velocity $u_*$ generated under the water surface
by the wind stress was provided in \cite{bauer83},
\beq
u_*\approx A[\hat\rho_{air}(1+u_{wind}/\bar u)]^{1/2}u_{wind},
\eeq
where $\hat\rho_{air}\approx 10^{-3}$ is the air water density ratio, and 
$A$ and $\bar u$ are
empirical constants: $A=0.028$ and $\bar u=12.3$ m/s. A 10-meter wind velocity $u_{wind}=10$ m/s
would lead to a friction velocity
\beq
u_*\approx 0.01\ {\rm m/s}.
\label{ustar}
\eeq
The strength of the convective forcing can be estimated from the heat flux to the
atmosphere. Estimates of the heat flux during frazil ice production events fall in 
the range
\beq
\Phi^{tot}_T=2.5\cdot 10^{-5} - 10^{-4}\ {\rm ^oC\,m/s},
\eeq
corresponding in energy units to $\rho_wc_P\Phi^{tot}_T=
100 - 1000\,{\rm W/m^2}$
\cite{bauer83,omstedt84,matsumura15}.
From Eqs. (\ref{ustar}) and (\ref{Phi_T^tot}) we can define an Obukhov depth signaling
transition from a mechanical stress dominated to a thermal convection dominated turbulent
region
\beq
L_T=\frac{u_*^3}{\alpha_T\Phi_T^{tot}}= 10 - 100\ {\rm m}
\label{obukhov T}
\eeq
(this is clearly a lower bound since the heat flux responsible for convection is only
a fraction of $\Phi_T$).
The Obukhov depth is going to be reduced by salt release in ice formation. If all the
heat ceded to the atmosphere came from ice formation we would get
\beq
L_S=
\frac{\hat\latent u_*^3}{\alpha_S\hat S_R\Phi_T^{tot}}
=1- 10\,{\rm m},
\label{obukhov}
\eeq
which tells us that already one tenth of $\Phi_T^{tot}$ coming from ice formation would
be sufficient to invalidate Eq. (\ref{obukhov T}).
The estimate in Eq. (\ref{obukhov}) would be further reduced if ice 
were transported down the column together with the brine,
\beq
L_C=\frac{\hat\latent u_*^3}{\alpha_C\Phi_T^{tot}}=0.3- 3\,{\rm m}.
\label{Obukhov C}
\eeq
Note, from the second of Eq.
(\ref{relative buoyancy}), that transport down the column of just 1/4 of the
ice produced at the surface, would be sufficient to counterbalance the destabilizing
effect of salinity production. 
\section{The turbulent boundary layer}
\label{The mechanical}
We assume that a well developed mechanical boundary layer, in which
the feedback by $C$, $T_o$ and $T_n$  on the turbulence can be disregarded, exists. 
We focus on a low entrainment regime, $\Entrainment\ll 1$, such that
the depth of the boundary layer is correctly estimated by $L_S$.
The friction velocity $u_*$ and the Obukhov length $L_S$ provide the natural
scales for the velocity fluctuations in that region.  
A natural scale for the reacting fields $C$ and $T_o$ is
the supercooling flux in the ice-free region, $\Phi^B_{T_o}\equiv\Phi_\calx$.

We rescale quantities in terms of $L_S$, $u_*$ and $\Phi_\calx$: 
\beq
&&z\to L_Sz,
\quad
t\to (L_S/u_*)t,
\nonumber
\\
&&T_{o,n}\to(\Phi_\calx/u_*)T_{o,n},
\nonumber
\\
&&
C\to
[u_*(\hat\latent+a_S\hat S_R)]^{-1}
\Phi_\calx C.
\label{adimensional}
\eeq
After rescaling, $\calx=T_o-C$. It is convenient to introduce the reacting field
\beq
\caly=T_o+C.
\label{caly}
\eeq
The ice dynamics is simplified considering that
only crystals with $u_R\ll u_*$ are transported down the column
(we are not interested here in the determination of the deposition fluxes at the surface, which 
would require studying the dynamics of the crystal size spectrum).
We thus neglect $u_r$ in Eq. (\ref{C equation}).
Transport in a 
horizontally homogeneous 
mechanical boundary layer can be modeled by introducing an
eddy diffusivity $\kappa_{turb}\approx -\sigma u_*z$, where
$\sigma=0.4$ is the  von Karman constant
(we neglect all sources of inhomogeneity, such as Langmuir turbulence \cite{leibovich83}).
Transport equations in dimensionless form 
for the fields $T_n$, $\calx$ and $\caly$ can be obtained
from Eqs. (\ref{T equation}-\ref{C equation})
by replacing the advection terms with an eddy diffusion.
Exploiting Eqs. (\ref{grande1}), (\ref{Bowen}) 
and (\ref{obukhov}) allows to eliminate any explicit dependence on the heat fluxes. We get
\beq
&&\partial_z(z\partial_z\bar\caly)+(1/2)[\overline{\lambda\caly^2}-\overline{\lambda\calx^2}]=0,
\label{mechanical1}
\\
&&\partial_z(z\partial_z\bar T_n)=\partial_z(z\partial_z\bar\calx)=0,
\label{mechanical2}
\eeq
where
\beq
\lambda=
\frac{H_\Bowen u_*\gamma\hat\latent}{\sigma\alpha_S\hat S_R},
\label{lambda}
\eeq
with $H_\Bowen=(\Bowen-\Bowen_{min})/(1-\Bowen_{min})$,
gives the relative strength of the contributions 
to the dynamics  from ice formation and transport. 
The parameter $\lambda$
could equivalently be seen as the  ratio of 
the turbulent transport timescale $L_S/u_*$ and
the reaction timescale $u_*/(\Phi_{T_o}^B\gamma)$,
which explains the counter-intuitive proportionality with $u_*$. 
It is interesting to note that $\lambda$ vanishes in the limit $\Bowen\to\Bowen_{min}$, 
corresponding to the limit $\Entrainment\to 0$ in Eq. (\ref{refined bound}).
Note that we allow in Eq. (\ref{mechanical1}) for the possibility of fluctuations
in $\lambda$, which would develop in the case of dependence of $\gamma$ 
on the sign of $T_o$ (see discussion at the end of Sec. \ref{Transport}).

We can make some estimates.
Assume turbulence strength $u_*\approx 0.01\ {\rm m/s}$ and 
one-mm crystals with $\epsilon= 1/50$.  The two estimates
$\gamma\approx 0.0037\ {\rm (^oC\, s)^{-1}}$ \cite{fujioka74,rees-jones15} and
$\gamma\approx 0.04\ {\rm (^oC\, s)^{-1}}$ \cite{hammar95} give 
$\lambda\approx 3 H_\Bowen$ and
$\lambda\approx 0.3 H_\Bowen$, respectively. Smaller crystals lead to larger values
of $\lambda$, but the effect is counteracted, at least in part, by the fact that such crystals 
typically have larger $\epsilon$ \cite{ghobrial13}. In general, $\lambda$ is
not small and the nonlinearity in Eq. (\ref{mechanical1}) cannot {\it a priori} be disregarded.

We see from Eq. (\ref{mechanical2}) that the
two non-reacting fields $T_n$ and $\calx$ obey logarithmic scaling
\beq
\bar T_n(z)=\frac{1}{\sigma}\frac{1-\Bowen_{min}}{\Bowen-\Bowen_{min}}\ln(-z/\Lambda_n),
\label{T_n(z)}
\eeq
and
\beq
\bar\calx(z)=(1/\sigma)\ln(-z/\Lambda_\calx),
\label{calx}
\eeq
where the factors in front of the logarithms are $(1/\sigma)$ times the fluxes $\Phi_{T_n}$ and
$\Phi_\calx$ expressed in dimensionless form (see Eqs. (\ref{Budget0}), 
(\ref{K_Bowen}) and (\ref{adimensional})), 
and 
$\Lambda_{n,\calx}$ are free parameters determined by the asymptotic large depth behavior
of the fields $\bar T$ and $\bar S$ (and $\bar C$ if it reaches the bottom of the layer). 

The two limit regimes $\Lambda_n\ll 1$ and $\Lambda_n\gtrsim 1$,
correspond to $\bar T_n>0$ and $\bar T_n<0$, respectively,
in most of the domain. 
In the same way, $\Lambda_\calx\ll 1$ and $\Lambda_\calx\gtrsim 1$ correspond to
$\bar\calx>0$ and $\bar\calx<0$ in most of the domain.
For $\Lambda_\calx\ll 1$, from positive definiteness of $C$,
most of the domain will be above freezing; for $\Lambda_\calx\gtrsim 1$, either
supercooling, or ice, or both will be present.

\subsection{Two limit regimes}
\label{Two limit}
For small $\lambda$, the dynamics of $\caly$ in the mechanical boundary layer reduces to that
of a passive scalar; $\caly$ and therefore also $T_o$ and $C$ obey in the first
approximation logarithmic scaling.

For $\lambda\gtrsim 1$, solution of Eq. (\ref{mechanical1}) 
is complicated by non-linearity and by the
presence of fluctuations. 
To make analytical progress, we momentarily neglect 
fluctuations.
We can linearize the dynamics when either $\bar C$ or $\bar T_o$ are small. 

Small $\bar C$ corresponds to a regime of small entrainment on the scale of $T_o$
and above freezing conditions: $\Phi_C^{top}\ll 1$, $\bar T_o>0$. In this
case, $\bar T_o\simeq\bar\calx$ and the domain of interest is $\Lambda_\calx\ll -z$.
This corresponds to studying the melting dynamics of $\bar C$ for fixed $\bar T_o>0$.

Small $\bar T_o$ corresponds to a regime in which large amounts
of frazil ice are present in  the column. In this case $\bar C\simeq-\bar\calx$ and 
the domain of interest is $-z\ll\Lambda_\calx$ (note that the frazil ice may
reach the bottom of the mechanical boundary layer). 
This corresponds to studying the decay of $\bar T_o$ induced by ice formation or melting
for fixed $\bar C$.

For small $\bar C$ we can approximate
$\bar\caly^2-\bar\calx^2=4\bar C\bar T_o \simeq 4\bar\calx\bar C$
and Eq. (\ref{calx}) becomes, exploiting Eqs. (\ref{mechanical2}) and (\ref{calx}),
\beq
\sigma\partial_z(z\partial_z\bar C)
+\bar \lambda\bar C\ln(-z/\Lambda_\calx)=0.
\label{mechanical3}
\eeq
For small $\bar T_o$ we can approximate
$\bar\caly^2-\bar\calx^2=4\bar C\bar T_o \simeq -4\bar\calx\bar T_o$
and Eq. (\ref{calx}) becomes, exploiting again Eqs. (\ref{mechanical2}) and (\ref{calx}),
\beq
\sigma\partial_z(z\partial_z\bar T_o)
-
\bar \lambda\bar T_o\ln(-z/\Lambda_\calx)=0.
\label{mechanical4}
\eeq
The logarithm on the right-hand side of Eq. (\ref{mechanical3}) gives the profile of $\bar T_o$, 
with $-z=\Lambda_\calx$ the depth of the supercooled region; in Eq. (\ref{mechanical4}),
the logarithm gives the profile of $-\bar C$, with $-z=\Lambda_\calx$
the maximum depth reached by the frazil ice. 

Equations (\ref{mechanical3}) and (\ref{mechanical4}) are identical in form
and describe decay (within logarithms) at depth $-z\sim \bar \lambda^{-1}$
of the respective field (note the minus sign in Eq. (\ref{mechanical4}), which cancels the negative
sign of the logarithm in the region $-z<\Lambda_\calx$). 
Let us calculate the decay explicitly.
We assume that decay takes place in the region of applicability of Eqs. (\ref{mechanical3})
and (\ref{mechanical4}), that is the mechanical boundary layer $-z<1$.

Consider first the case of Eq. (\ref{mechanical3}),
in which $\bar C$ is small and $\bar T_o\simeq\bar\calx$.
The singularity at $z\to-\infty$ in Eq. (\ref{mechanical3}) is treated by
working in the eikonal representation, $\bar C=\exp(W)$ \cite{bender}.
To leading order in $z$, Eq. (\ref{mechanical3}) becomes
\beq
\sigma z(W')^2+\bar\lambda\ln(-z/\Lambda_\calx)=0,
\eeq 
which has solution, for $-z\gg\Lambda_\calx$,
\beq
W(z)=\pm 2\sqrt{(
-z\bar\lambda/\sigma
)\ln(-z/\Lambda_\calx)}.
\eeq
We get the asymptotic behavior
\beq
\bar C(z)\sim\exp\left[-2\sqrt{(-z\bar\lambda/\sigma)
\ln(-z/\Lambda_\calx)}\right],
\eeq
which leads to the decay depth for $\bar C(z)$
\beq
-z_C\sim |\bar\lambda\ln(\bar\lambda\Lambda_\calx)|^{-1}.
\label{z_C}
\eeq

The case in which $\bar T_o$ is small and $\bar C\simeq-\bar\calx$ is treated in exactly
the same way. We proceed with Eq. (\ref{mechanical4}) as with Eq. (\ref{mechanical3}),
except that now $-z\ll\Lambda_\calx$.
 We obtain for $\bar T_o(z)$ the same decay law 
in Eq. (\ref{z_C}), $-z_{T_o}\sim
|\bar\lambda\ln(\bar\lambda\Lambda_\calx)|^{-1}$.

\section{Fluctuations}
\label{Fluctuations}
\begin{figure}
\begin{center}
\includegraphics[width=0.6\columnwidth]{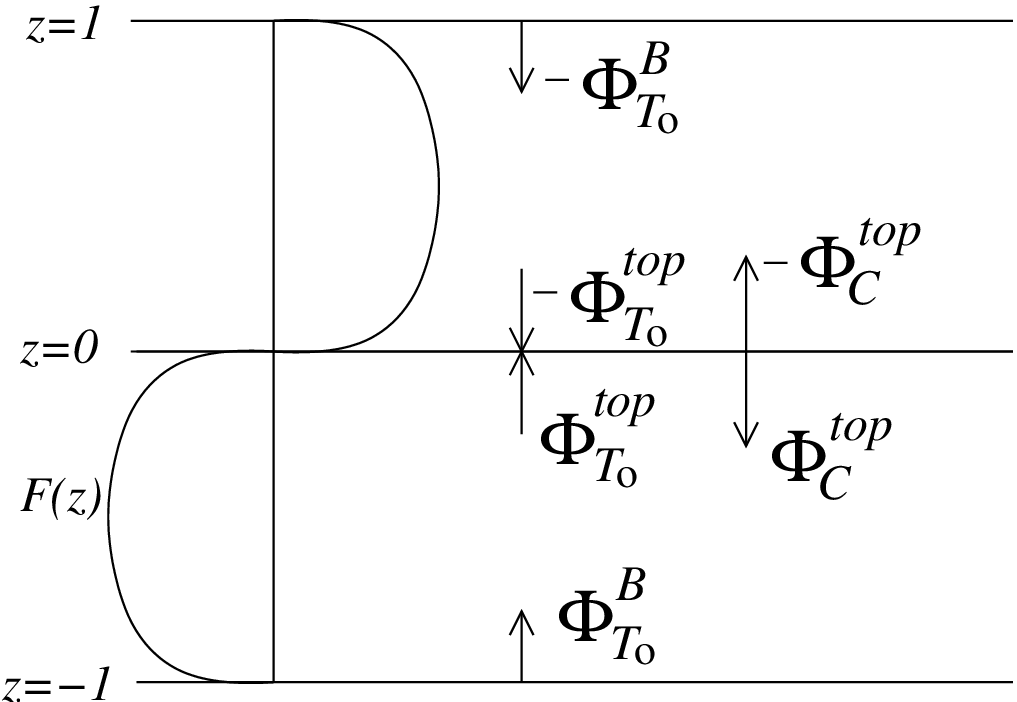}
\caption{Sketch of the computation domain.
}
\label{schema1}
\end{center}
\end{figure}

We estimate fluctuations in the mechanical boundary layer by means of a simple
Kraichnan model \cite{kraichnan68,falkovich01},
in a periodic 2D domain $[-1,1]\times [-1,1]$. We focus on the dynamics
of the reacting fields $T_o$ and $C$. We take $x$ and $z$ as the horizontal and vertical
coordinates, with $z=0$ the water surface and the periodic point $z=\pm 1$
the bottom of the boundary layer ($z=-L_S$ in the original units). The computation domain
is thus split in two statistically equivalent mirror sub-domains at $-1<z<0$ and $0<z<1$.
The situation is illustrated in Fig. \ref{schema1}.
We continue to work with the dimensionless variables  defined in 
Eq. (\ref{adimensional}).

A boundary layer structure
is imposed on the velocity fluctuations by means of a
shape function $F=F(z)$. We write
\beq
u_x=\bar u_x-\partial_z\psi_F,
\qquad
u_z=\partial_x\psi_F,
\label{uxz}
\eeq 
where
$\psi_F(\r,t)=F(z)\psi(\r,t)$ and
$\psi(\r,t)$ is zero-mean, spatially homogeneous, and white in time,
\beq
\overline{\psi(\r,t)\psi(\r',t')}=\Psi(\r-\r')\delta(t-t').
\label{Kraichnan psi}
\eeq
We take for the shape function
\beq
F=\prod_{m\in\mathbb{Z}}\tanh[2\pi(z+m)]
\label{shape function}
\eeq
and for $\bar u_x(z)$ a sum of logarithms mimicking the mean velocity profile in 
a turbulent channel flow.

We chose the shape function in such a way that
$u_z(0)=0$ but $u_x(0)\ne 0$, as
expected for the turbulent velocity at the free water surface. The fact that 
$F'(0)\ne 0$, together with periodicity, require however that
$F=0$ and therefore $u_z=0$ somewhere else in the domain. The choice $F(\pm 1)=0$
is the one that less affects the dynamics, although it somewhat spoils
the interpretation of $z=\pm 1$ as the bottom of the mechanical boundary layer.

We take for the spectrum $\Psi_\k=\int\d^2r\ \ex^{-\im\k\cdot\r}\Psi(\r)$,
\beq
\Psi_\k=A\ (k^2+k_0^2)^{-8/3},
\eeq
which guarantees Richardson scaling for relative diffusion at 
small separation \cite{falkovich01}. 
The parameter $k_0$ is a large scale cutoff that we put equal
to $5\pi$.
The constant $A$ is fixed by imposing the condition for the spatial
average of the turbulent
diffusivity
\beq
\langle\kappa_{turb}\rangle=\int\d t\  \langle\overline{u_z(\r,t)u_z(\r,0)}\rangle=1,
\eeq
which replaces the normalization $u_*=1$ implicit in Eq. (\ref{adimensional}). 

The presence of input and output fluxes for $C$ and $T_o$ at the boundaries of
a periodic domain is
mimicked introducing forcing terms at $z=0$ and $z=-1$ in the transport equations
(see Fig. \ref{schema1}).
From Eqs. (\ref{T equation}-\ref{C equation}) and (\ref{adimensional}), we find
(with $z$ mod. 2):
\beq
&&(\partial_t+\u\cdot\nabla)T_o=
\kappa\nabla^2T_o
-\lambda CT_o
\nonumber
\\
&&+2[-(\Phi_C^{top}+1)\delta(z)+\delta(z+1)],
\label{Kraichnan To}
\\
&&[\partial_t+(\u+\u_r)\cdot\nabla]C=
\kappa\nabla^2C
-\lambda CT_o
\nonumber
\\
&&-2\Phi_C^{top}\delta(z),
\label{Kraichnan C}
\eeq
where $\kappa$ is understood as the turbulent diffusivity of the unresolved eddies 
below the spatial discretization scale 
and we recall, in dimensionless units,
$\Phi_{T_o}^B=1$, $\Phi_{T_o}^{top}=1+\Phi_C^{top}$.

Taking the difference of Eqs. (\ref{Kraichnan To}) and (\ref{Kraichnan C}), we see
that the spatial average of $\calx=T_o-C$ is conserved,
$(\d/\d t)\langle\calx\rangle=0$, which is equivalent to the statement on conservation of
$\Phi_\calx$ in the previous section. 

\begin{figure}
\begin{center}
\begin{minipage}[]{0.8\columnwidth}
\includegraphics[width=\columnwidth]{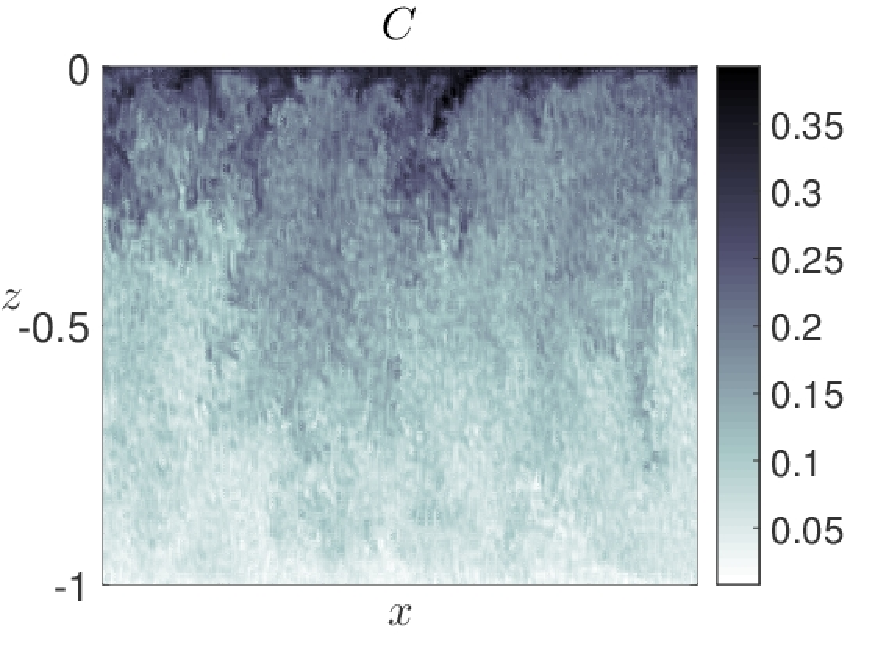}
\end{minipage}
\begin{minipage}[]{0.8\columnwidth}
\includegraphics[width=\columnwidth]{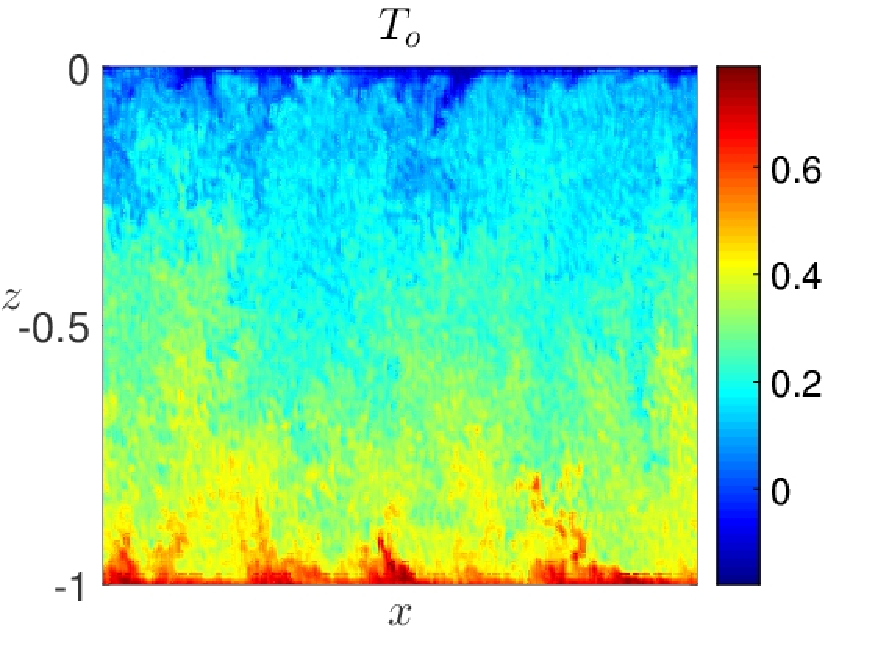}
\end{minipage}
\caption{Snapshot of the ice concentration (top panel) and the supercooling (bottom panel)
for $\langle\calx\rangle=-\Phi_C^{top}=0.1$, $\lambda_-=1$ and $\epsilon=1/5$.}
\label{splot}
\end{center}
\end{figure}

We solve numerically the system of equations 
(\ref{Kraichnan To}-\ref{Kraichnan C}), by means of a pseudospectral code 
on a $256\times 256$ grid, taking $\kappa=10^{-3}$. 
We smooth the white noise in Eq. (\ref{Kraichnan psi}) by replacing the Fourier modes 
$\psi_\k(t)$ with an
Ornstein-Uhlenbeck processes with correlation time below the diffusion time 
at the discretization scale.
An Adam-Bashford integration scheme has been used for 
advancement in time \cite{elhmaidi93,elhmaidi05}. 
A snapshot of the fields $C$ and $T_o$ is shown Fig. \ref{splot}.

\begin{figure}
\begin{center}
\begin{minipage}[]{0.65\columnwidth}
\includegraphics[width=\columnwidth]{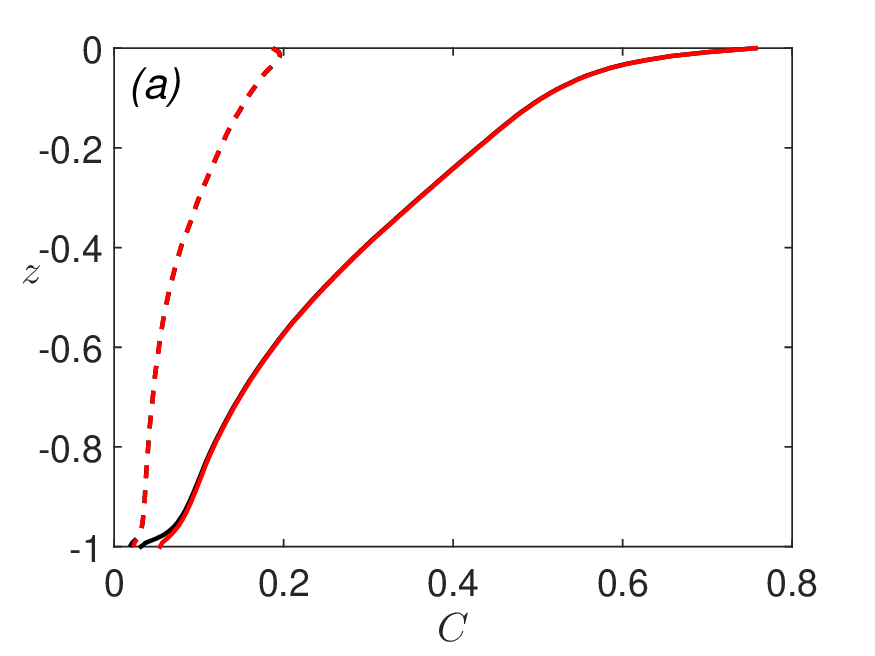}
\end{minipage}
\begin{minipage}[]{0.65\columnwidth}
\includegraphics[width=\columnwidth]{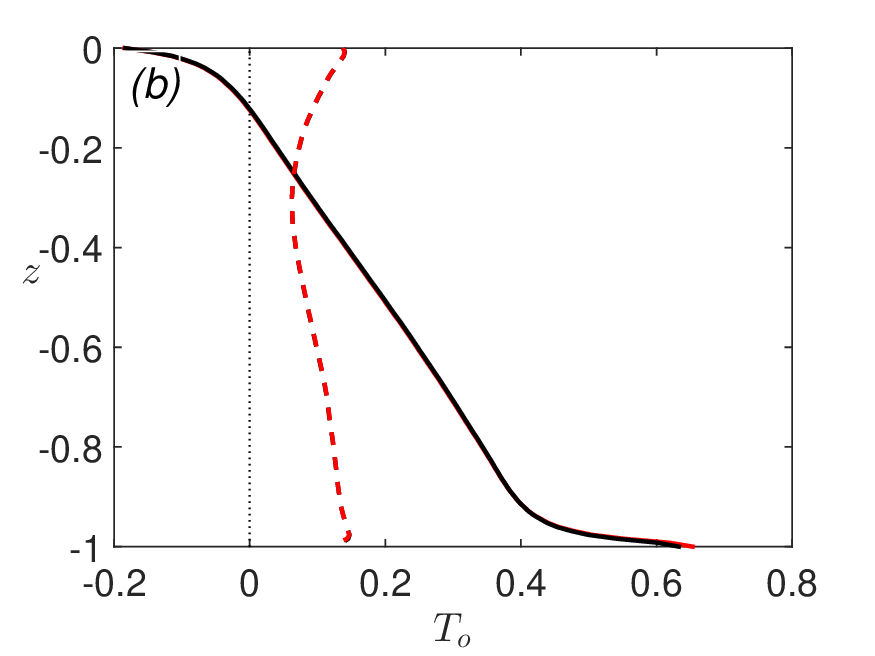}
\end{minipage}
\begin{minipage}[]{0.65\columnwidth}
\includegraphics[width=\columnwidth]{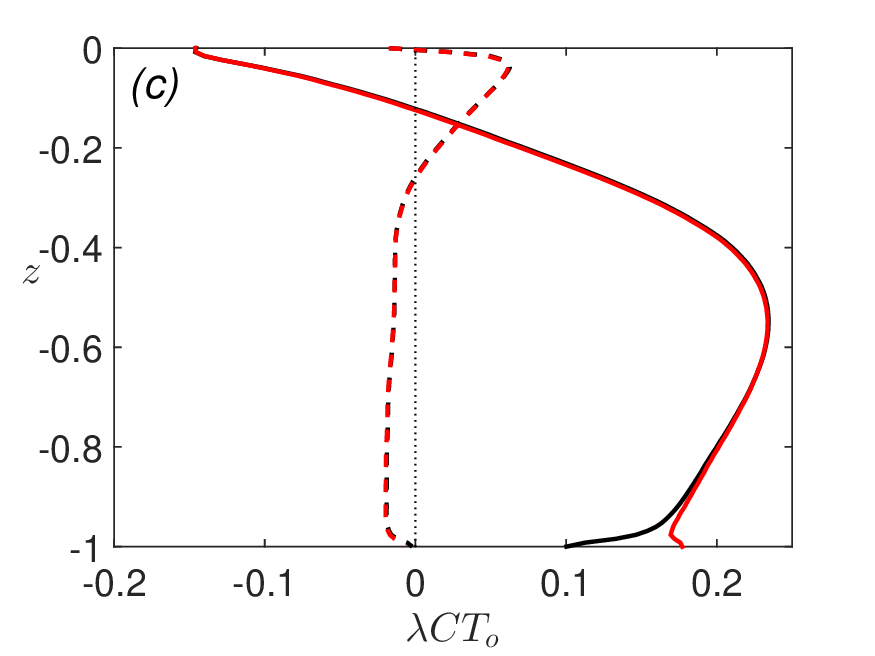}
\end{minipage}
\caption{
Effect of an ice  sink at $z=\pm 1$ on the 
vertical profiles of ($a$) ice concentration, ($b$) supercooling
and ($c$) ice production rate;
$\Phi_C^B=0$ in red; $\Phi_C^B=-0.3$ in black. 
In $a$ and $b$, solid lines indicate
average; in $c$, they indicate mean field result $\bar\lambda\bar C\bar T_o$.
In $a$ and $b$, dashed line indicate rms; in $c$, they indicate the
fluctuation $\overline{(\lambda CT_o)_f}$.
Values of 
other parameters: $\lambda_-=1$; $\Phi_C^{top}=-0.5$; $\langle\calx\rangle=-0.1$;
$\epsilon=1/5$. 
}
\label{profile0}
\end{center}
\end{figure}
\begin{figure}
\begin{center}
\includegraphics[width=0.65\columnwidth]{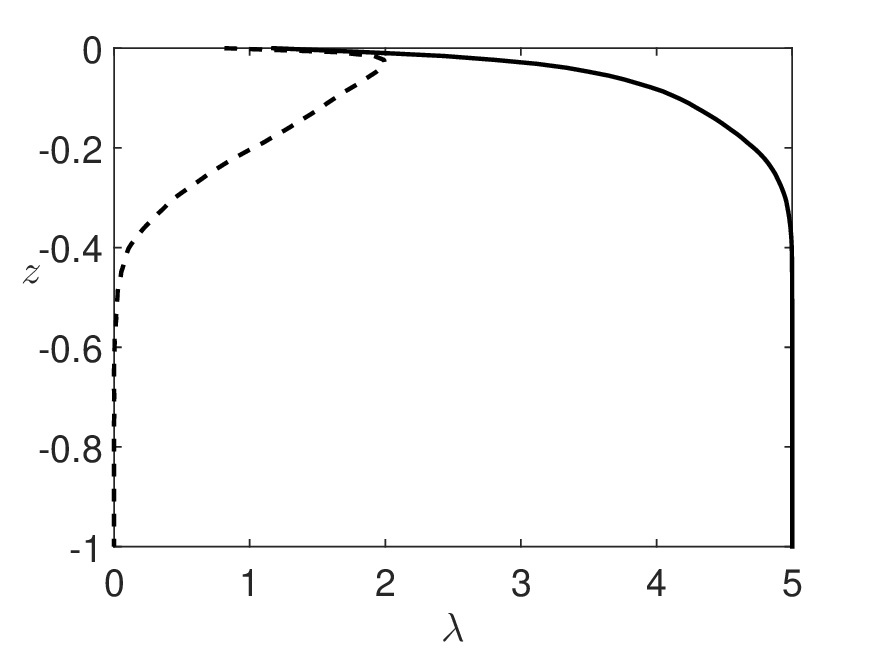}
\caption{Vertical profile of $\lambda$. Average solid; rms dashed.
Values of the parameters $\lambda_-=1$; $-\Phi_C^{top}=\langle\calx\rangle=0.1$;
$\epsilon=1/5$.
}
\label{lambda_1}
\end{center}
\end{figure}
\begin{figure}
\begin{center}
\includegraphics[width=0.65\columnwidth]{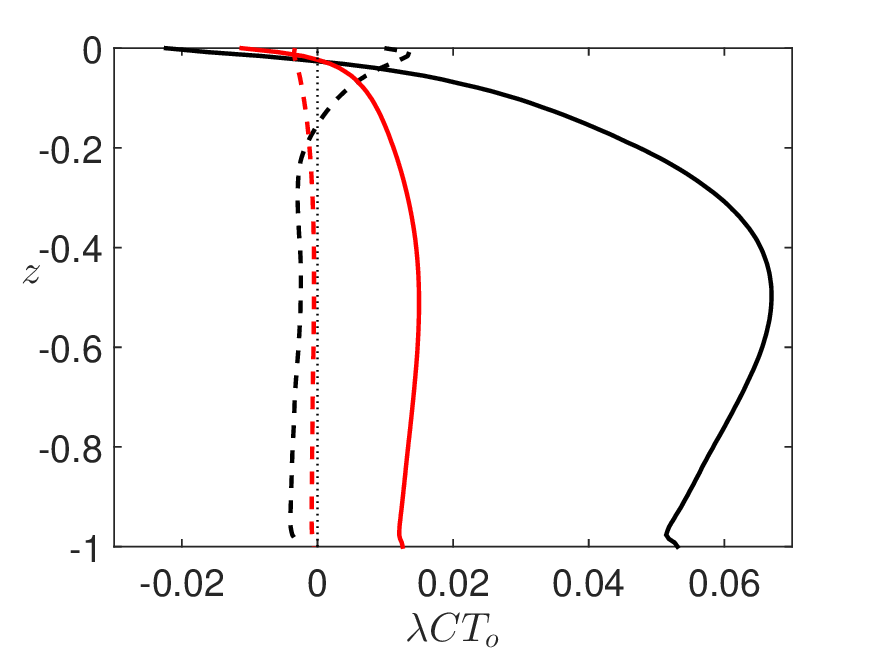}
\caption{
Vertical profiles of the ice production for different values of $\epsilon$.
$\epsilon=1$ in red, $\epsilon=1/5$ in black; mean field contribution $\bar\lambda\bar C\bar T_o$
solid line, fluctuation contribution $\overline{(\lambda C T_o)_f}$ dashed line.
Values of the other parameters: $\lambda_-=-\Phi_C=\langle\calx\rangle=0.1$.
}
\label{prod_01}
\end{center}
\end{figure}
\begin{figure}
\begin{center}
\includegraphics[width=0.65\columnwidth]{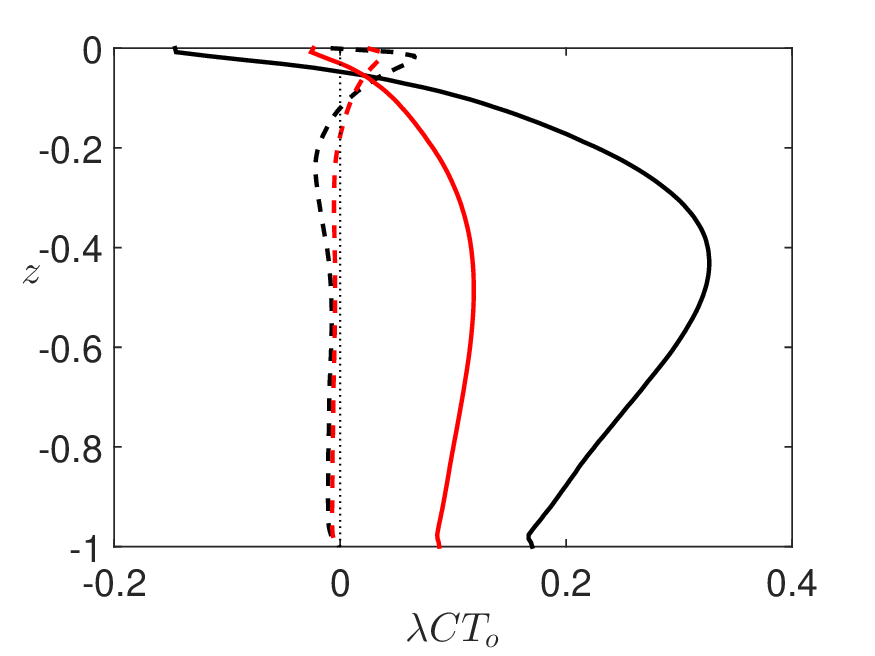}
\caption{
Same as Fig. \ref{prod_01} for
different values of $\epsilon$ and $\lambda$.
$\lambda_-=0.02$, $\epsilon=1/50$ in red; $\lambda_-=1$, $\epsilon=1/5$ in black.
}
\label{prod_e50_e5.eps}
\end{center}
\end{figure}
The production term $\lambda$ is defined following \cite{jenkins95,holland05}, 
\beq
\lambda_-\equiv\lambda_{T_o<0}=\epsilon^{-1}\lambda_{T_o>0}.
\label{lambdameno}
\eeq
One of the motivations for this choice is that Eq. (\ref{lambdameno}) 
provides a source of fluctuations analogous to those which 
could be expected, presumably, 
from taking into account the finite size spectrum of the crystals. 

As control parameters for the simulations we take
$\lambda_-$, $\epsilon$, $\Phi_C^{top}$ and $\langle\calx\rangle$.
We concentrate on the three cases $\lambda_-=1$, $\lambda_-=0.1$ and 
$\lambda_-=0.02$, which would correspond for $u_*=0.01\ $m/s to crystal 
radii $R\approx 0.14\ $mm, $R\approx 0.45\ $mm and $R\approx 1\ $mm 
\cite{note}.
We take $\epsilon=1/50$ for the largest crystal and $\epsilon=1/5$ for 
the smaller ones, in accordance with the  observations in \cite{ghobrial13}. 
The case $\epsilon=1$ is also considered to evaluate the contribution to fluctuations
from sources others than the 
growth-melt
asymmetry in Eq. (\ref{lambdameno}).

We put $\bar u_x(z)=0$, as inclusion 
of a non-zero horizontal mean velocity profile has been observed to produce 
negligible effects on the dynamics. As done in Sec. \ref{The mechanical}, we put $\u_r=0$,
which is appropriate for the smallest crystals, but may be a rough approximation for 
the largest ones (see Fig. \ref{frazfig1}).

We note that ice may be present at the bottom of the domain, $z=\pm 1$, thus
generating spurious ice fluxes.  To evaluate the
effect  on the dynamics, we compare with the case in which
an ice sink $\Phi_C^B$ is artificially added at $z=\pm 1$,
with $\Phi_{T_o}^B\to\Phi_{T_o}^B+\Phi_C^B$, to
guarantee conservation of $\langle\calx\rangle$ (with a slight abuse of notation we are 
using superscript $B$ for the bottom of the numerical domain although the region is not ice-free).
As shown in Fig. \ref{profile0}, the dynamics
is modified only in the boundary layer region near $z=\pm 1$, the
curves being essentially indistinguishable in the rest of the domain.
We thus expect that the dynamics
is properly taken into account by the model for generic values of the parameters.

As a general rule, we find that the fluctuation contribution to the production term 
$\overline{(\lambda CT_o)_f}=\overline{\lambda CT_o}-\bar\lambda\bar C\bar T_o$ is small,
even though the rms component of the individual fields $C$, $T_o$ and $\lambda$ are not
small at all. 
We see in Fig. \ref{profile0}($b$) and \ref{profile0}($c$), that
ice production fluctuations are concentrated around the transition region from negative
to positive $T_o$. We see in Fig. \ref{lambda_1} that fluctuations in $\lambda$ concentrate 
roughly in the same region, which
suggests that a strong contribution to the ice production fluctuations comes from
the difference in melting and freezing rates described in Eq. (\ref{gamma}).

To determine the effect of the fluctuations of $\lambda$ on the dynamics,
we compare the case of frazil crystals with $\epsilon=1$ and $\epsilon=1/5$.
We do not consider the contribution to fluctuations from 
the dependence of $\lambda$ on the crystal distribution, which remains undetermined
in a description based on the integrated field $C(\r,t)$. We observe in Fig.
\ref{prod_01} that for $\epsilon=1$, $\overline{(\lambda C T_o)_f}
=\lambda\overline{\tilde{C}\tilde{T_o}}<0$, 
suggesting a picture in which fluctuations of $CT_o$ arise 
from cold water parcels rich in ice transported by turbulence 
from $z=0$ into the body of the domain.
As shown in Figs. \ref{profile0}($c$), \ref{prod_01} and \ref{prod_e50_e5.eps}, for $\epsilon<1$,
fluctuations and mean field components give contribution to ice production of opposite sign 
in most of the domain.

Fluctuations become negligible deeper in the the column, where melting is dominant. 
This tells us that the mean field analysis in the previous section may be
appropriate for the decay of $C$ (see Eq. (\ref{z_C})), but not for that of $\bar T_o$,
due to the fluctuations of $\lambda$
in the regions where $\bar T_o\to 0$ and therefore $T_o$ undergoes most changes of sign.

\section{Conclusion}
We have studied the process of ice formation in a turbulent, horizontally homogeneous stationary 
water column, as a balance of fluxes of heat, salinity and ice. The following are our main 
results.

Imposing supercooling
decrease with depth has allowed us to derive lower bounds on the ratio of sensible
and latent heat fluxes to the atmosphere, which contain important information on the 
process of ice formation in the water column.

The minimum sensible heat flux ranges from roughly 1/2 to a few
percents of the total flux depending on the strength of ice entrainment. 
The sensible heat in excess to the lower bound may be associated with
ice melting at the bottom of the layer, or with direct heat transfer to the atmosphere,
depending on whether a sufficiently thick grease ice layer is present or not at the 
water surface. 

In the presence of 
entrainment, the minimum sensible heat flux accounts for the heat from the bottom of the
column required to melt the entrained ice crystals.  The same amount of heat
is released at the surface 
during ice formation. Entrained frazil ice behaves like a sort of conveyor belt for the heat,
providing a contribution to the total flux, which, depending on the level of entrainment, may 
be comparable to the latent heat contribution from net ice production. 

We have derived mean-field analytical expressions 
for the vertical profiles of salinity, temperature 
and ice concentration in the limit of low entrainment, assuming the presence of a 
wind-induced mechanical boundary layer at the top of the water column.

It appears that, except in situations in which all of the water column
is well above freezing, and entrainment is so low that the entrained ice
completely melts, frazil ice fluxes reaching the bottom of the mechanical 
turbulent layer are present.
At the same time, supercooled conditions can be found
at substantial depth in the water column,  confirming observations
in the laboratory \cite{ushio93}, in field campaigns
\cite{ito15,skogseth09,dmitrenko10} and in numerical simulations \cite{matsumura15}.

We explain the symmetric possibility of supercooling and frazil ice presence deep in the column,
with the existence of a single `reacting' field
actually being affected by ice formation (the field $\caly$ in Eq. (\ref{caly})).
This has the consequence that the dynamics
of supercooling and ice concentration, in the two cases in which one of the fields
is small, are essentially identical.

We explain the observation that frazil ice and supercooling may be present 
at substantial depth in the water column with the smallness of the
growth or melting rate of the ice crystals compared to the rate of turbulent mixing
(the parameter $\lambda$ in Eq. (\ref{lambda})). 
The depths (normalized to the haline
Obukhov depth) of the supercooled region in the presence of large frazil ice amounts, 
and of the frazil ice region in above freezing conditions, are both $\sim\lambda^{-1}$.

We stress that all the above results on ice dynamics in the mechanical boundary
layer are conditioned to smallness of the ratio of the rate of entrainment of frazil ice 
to total ice production (the parameter $\Entrainment$ in Eq. (\ref{Entrainment})).
It is not clear whether our results would be confirmed in the presence of strong entrainment,
in which case,
turbulence stabilization by frazil ice may may lead to a double-convection regime.

The mean-field results are confirmed by numerical analysis by means of a 
two-dimensional Kraichnan model. The analysis shows that a strong contribution to 
fluctuation is produced when asymmetry between the rates of ice formation and melting
is assumed, as done in  \cite{smedsrud04,holland05};
otherwise, fluctuations are small. This suggests
that more accurate descriptions of ice formation, in which
the size spectrum of the crystals and the 
different growth and melting rates of the different size classes
are taken into account, could lead to similar higher fluctuation levels.

\vskip 10pt
\noindent
{\bf Acknowledgements}
\\
This research was supported by FP7 EU project ICE-ARC
(Grant agreement No. 603887), by MIUR-PNRA, PANACEA project (Grant No. 2013/AN2.02)
and by COST Action grant MP1305.





\begin{thebibliography}{00}
\bibitem{note00}
In order for a flux-gradient relation between $\Phi_{T_o}$ and $T_o$ to hold,
it is necessary that $T$ and $S$ are transported together, which is reasonable
if molecular diffusion can be disregarded.
\bibitem{note0}
We are assuming that the microscopic flows in the ice layer are produced 
by mechanical means, such as wave stirring and wind-induced shear motions.
Linearity in $T$ and $S$ could be violated by
convection at the crystal scale.
\bibitem{note}
We are assuming for $\gamma$ the parametrization in \cite{hammar95}. 
The alternative parametrization
in \cite{fujioka74,rees-jones15} would lead to values of $R$ larger by a factor 
$\approx 3.3$.
\end{thebibliography}

\section*{References}
\label{References}

\end{document}